# Uncertainty-Aware Machine-Learning Framework for Predicting Dislocation Plasticity and Stress–Strain Response in FCC Alloys

Jing Luo[1], Yejun Gu[2,1], Yanfei Wang[3], Xiaolong Ma[4], Jaafar.A El-Awady[1*]

[1*]Department of Mechanical Engineering,, Johns Hopkins University, 3400 N Charles Street, Baltimore, 21218, MD, USA.
[2]Institute of High Performance Computing, Agency for Science, Technology and Research, 138632, Singapore.
[3]School of Aerospace Engineering, Xi'an Jiaotong University, Xi'an,710049, China.
[4]Department of Materials Science and Engineering, City University of Hong Kong, Hong Kong, China.

*Corresponding author(s). E-mail(s): jelawady1@jhu.edu;

**Abstract**

Machine learning has significantly advanced the understanding and application of structural materials, with an increasing emphasis on integrating existing data and quantifying uncertainties in predictive modeling. This study presents a comprehensive methodology utilizing a mixed density network (MDN) model, trained on extensive experimental data from literature. This approach uniquely predicts the distribution of dislocation density, inferred as a latent variable, and the resulting stress distribution at the grain level. The incorporation of statistical parameters of those predicted distributions into a dislocation-mediated plasticity model allows for accurate stress–strain predictions with explicit uncertainty quantification. This strategy not only improves the accuracy and reliability of mechanical property predictions but also plays a vital role in optimizing alloy design, thereby facilitating the development of new materials in a rapidly evolving industry.

**Keywords:** Mixed density network, Dislocation, Stress-strain curves, FCC polycrystal materials, Uncertainties

# 1 Introduction

The design and discovery of high-performance alloys have long relied on labor-intensive "trial-and-error" research approaches that are both slow and expensive (cf. [1]). The challenge is further amplified by the large combinatorial space of compositions, processing routes, and microstructural states, together with the limited throughput of conventional experimentation. During the past decade, artificial intelligence (AI) and machine learning (ML) methods have emerged as powerful tools to navigate this vast design space[2, 3], allowing rapid screening of alloy chemistries and microstructures for targeted properties such as strength [4], ductility [5], or corrosion resistance [6]. However, despite the promise of these data-driven methods, one critical challenge remains: incorporating and quantifying the multiple layers of uncertainty inherent in both experiments and theoretical/computational models.

Uncertainty in materials research can be broadly classified as (i) aleatory uncertainty (i.e., random variability originating from intrinsically stochastic features such as microstructural characteristics) and (ii) epistemic uncertainty (i.e., lack of knowledge or missing information, for example due to small or incomplete data sets) [7]. Ignoring these uncertainties leads to overly simplistic models that are incorrectly "over-confident" and that generalize poorly when extrapolated to new alloys or processing histories.

A classical example of how uncertainty is manifested in mechanical property data is evident from the published stress-strain curves for any single metal (pure or alloy), as shown in Supplementary Figure S1(a). The published data show substantial scatter from one experiment to another, even when the reported nominal composition and microstructural features are nearly identical. In turn, empirical models for strength predictions, such as the well-known Hall-Petch relationship [8, 9], which describes how the average grain size correlates with the yield strength of a material, are prone to these uncertainties. Despite its success in capturing broad trends, published experimental reports on seemingly identical specimens often exhibit strength values that can differ by an order of magnitude, as shown in Supplementary Figure S2(a) through (c), which are not reflected by the average trends predicted by the Hall-Petch relationship. Such inconsistencies often arise from variations in microstructural characteristics that are closely related to specific processing methods, such as grain size distribution, grain orientation, average initial dislocation density, and local compositional variations[10]. However, the literature typically reports only their averaged values. Therefore, when models such as the Hall-Petch relationship are developed without accounting for these microstructural features and their associated uncertainties, it becomes extremely difficult to predict the wide scatter in experimental data or to extract reliable parameters that are transferable from one material to another.

To overcome the inherent averaging of such phenomenological relationships, several computational frameworks have been proposed to capture the microstructural effects on mechanical properties more rigorously. One of the most widely used methods is the crystal plasticity finite element method (CPFEM), which can incorporate grain orientations, slip systems, and other microstructural features to predict the stress-strain response of different alloys [11–15]. Over the years, CPFEM studies have offered insight into how microstructural features affect macroscopic behavior. However, CPFEM typically relies on constitutive phenomenological equations that must be calibrated for each new alloy or even each new experiment, and they are associated with a high computational cost, thereby limiting its generalizability for exploring large spaces of the design space. Beyond CPFEM, methods such as atomistic simulations [16–18] or discrete dislocation dynamics (DDD) simulations [19–22] can also incorporate aspects of local microstructures and quantify their effects on the mechanical properties of metals. Nevertheless, each of these methods have their own computational cost and modeling assumptions and may still require experimental input for parameter calibration.

Given the growing need for high-throughput uncertainty-aware predictions of mechanical behavior, multifidelity frameworks have recently been explored to couple data-driven methods with physics-based modeling [23, 23]. However, when predicting the stress-strain response of alloys, most existing approaches, if not all, fail to capture the broad variability introduced by dislocation-mediated plasticity, grain boundaries, chemical inhomogeneities, and measurement inconsistencies, all of which are interrelated and collectively exacerbate the associated uncertainties. This gap between modeling and reality will persist unless these influences are carefully quantified.

Here, we present an integrated methodology that addresses this challenge by (1) explicitly incorporating both aleatory and epistemic uncertainties into an ML-based framework, (2) leveraging a statistical perspective to capture all possible microstructural variabilities and their effect on stress-strain behavior, and (3) demonstrating how the approach can be extended to a broad range of face-centered cubic (FCC) alloys without recalibration.

The remainder of this paper is organized as follows. Section 2 outlines the key challenges in predicting the stress-strain response in metals and discusses the limitations of existing models, and then introduces our uncertainty-aware machine learning framework to predict the evolution of dislocation density as a function of strain. Section 4 presents the performance of the model in predicting both the evolution of dislocation density and the stress-strain response of FCC metals, including its extension to alloys beyond the training dataset. Section 5 then investigates the sources of uncertainty in the predictions of the stress-strain curve and the underlying mechanisms that support the transferability of the model to other alloys. Finally, Section 3 provides a detailed description of the methods used throughout the study.

# 2 Key Challenges in Predicting the Stress-Strain Response of FCC Metals

Accurate prediction of the stress-strain response of FCC metals, in which dislocation plasticity is the key mechanism of deformation, requires several essential components. First, at the grain level: (i) a constitutive law relating grain stress to the dislocation density in the grain, $\rho$, and (ii) an evolution law that captures how $\rho$ changes with strain, $\epsilon$. Second, at the polycrystalline level: (iii) the microstructural heterogeneities (e.g., distributions of grain size/orientation and local compositional variations) that distinguish real materials and (iv) the grain-level models must be combined through an appropriate homogenization framework.

The generalized size-dependent Taylor strengthening (GS-DTS) law for a single grain can be described as [24]:

$$\sigma = \frac{1}{M}\left(\tau_0 + \alpha\mu b\sqrt{\rho} + \frac{\beta\mu}{d\sqrt{\rho}}\right), \tag{1}$$

where $M$ is the Taylor (or orientation) factor, $\mu$ is the shear modulus, $b$ is the Burgers vector, $\tau_0$ is the critical resolved shear stress for dislocation slip in a grain having size $d$, and $\alpha$ and $\beta$ are constants that depend on the crystal structure.

Next, the evolution of the dislocation density, $\rho$, with $\epsilon$ must be specified to predict the stress-strain evolution within the grain. Several phenomenological models have previously been proposed in the literature for this purpose [25–35], among which the Kocks–Mecking–Estrin (KME) model [29–35] has been widely used. This model can be expressed as:

$$\frac{d\rho}{d\epsilon} = K_0 d^{-1} + K_1 \rho^{1/2} - K_2 \rho, \tag{2}$$

where $K_0$, $K_1$, and $K_2$ are fitting parameters determined from experimental results. Although the KME model and other phenomenological models have been widely used in both analytical (e.g., [36]) and computational (e.g., [37, 38]) studies, the large scatter in the experimental data undermines their reliability. This is evident from the analysis discussed in Supplementary Materials Section S1. Although the KME model performs well when carefully calibrated for individual experiments as shown in Supplementary Figure S1(b), when attempting to fit the parameters to multiple experiments for the same material simultaneously, the fitting parameters $K_0$, $K_1$, and $K_2$ vary dramatically when fitted across multiple experiments on the same material. This is true even for pure metals, as shown in Figure S1(c) and Figures S3(a) through (c) for pure Ni. Similar scatter is observed for Cu and Al, as well as when using other phenomenological models (e.g., [26]) are used. This can be attributed to the fact that these models are phenomenological with parameters that are calibrated based on the average grain size and the average dislocation density, overlooking other important microstructural details such as the grain orientation, grain size, and dislocation density distributions. This highlights the insensitivity of these models to the variabilities inherent in experiments, which is a fundamental limitation of the use of such phenomenological approaches in exploring a wide range of chemistries and microstructures.

In addition, the experimental full three-dimensional (3D) microstructure characterization of polycrystals, initial dislocation density measurements, and in situ investigations of the evolution of the dislocation density are tedious and resource intensive [39], often relying on expensive instrumentation or specialized radiation sources (e.g., [40]). Given these challenges, experimental data sets in the literature are mostly incomplete or limited, making it nearly impossible to capture a fully representative understanding of the contribution of different microstructural features on the evolution of dislocation density. Consequently, deterministic models that require large parameters, such as the phenomenological KME model, are not suitable to robustly predict the evolution of the dislocation density as a function of strain in different compositions and microstructures without extensive recalibration.

Together, these challenges underscore the need for a more robust framework, one that inherently accounts for uncertainty in microstructures and captures the statistical nature of dislocation evolution across a broad range of conditions. Such an approach would not only provide more reliable predictions of the stress-strain response but would also guide the development of advanced materials by identifying and quantifying the factors that most strongly influence material performance.

Here, we develop a coupled physics-based and data-driven framework (see Fig. 1), to address the shortcomings of the conventional phenomenological models for predicting the stress-strain curves of FCC polycrystalline metals. The framework includes: (i) training a mixed density network (MDN) [41] on experimental data from the literature to capture the uncertainty in experimental results; (ii) a methodology to synthetically represent polycrystalline microstructures given the expected microstructural statistic parameters; and (iii) laws based on physics that connect the MDN predictions at the single grain level with the polycrystalline response.

## 3 Methods

For the MDN training, the inputs include the five most widely reported features, including true strain, $\epsilon$, strain rate, $\dot{\epsilon}$, average grain size, $d_{ave}$, shear modulus, $\mu$, and yield strength, $\sigma_\mathrm{y}$. The MDN output is the probability density function (PDF) of the grain dislocation density $\rho$ at each imposed strain. Returning a PDF rather than a single value, the model provides explicit confidence bounds that reflect both aleatory (microstructural) and epistemic (data) uncertainty in the experimental data. Details of the model training are given in Section 3.1, Section 3.4 and Section 3.5.

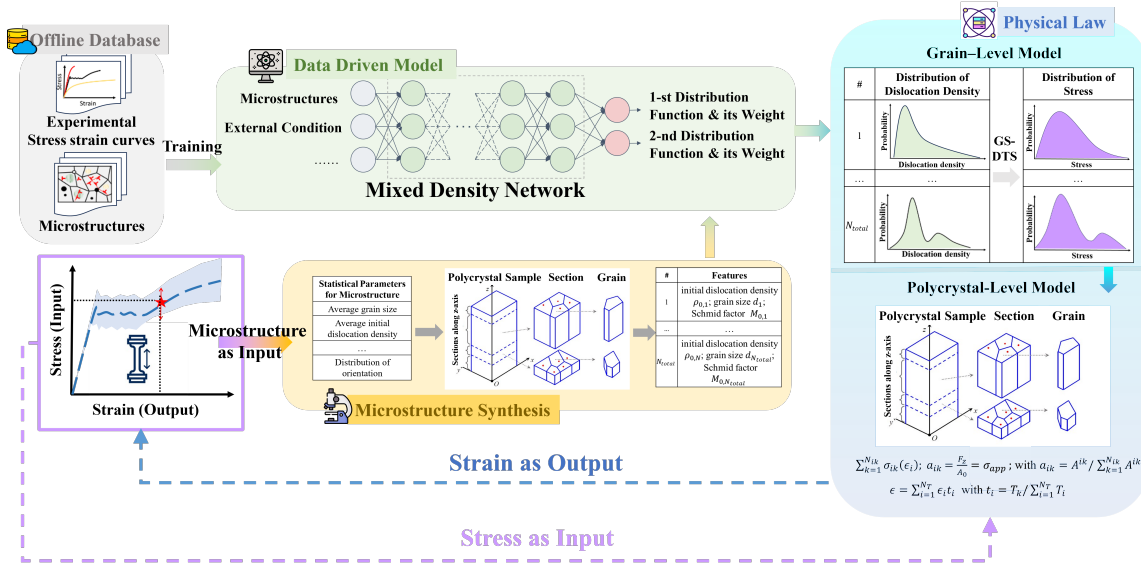

**Fig. 1** Flowchart illustrating the implementation of the coupled physics-based and data-driven framework for predicting the stress–strain curves of metals with uncertainty.

Once the MDN is trained, the stress–strain prediction for a new polycrystal proceeds through four sequential steps. First, a synthetic 3D rectangular polycrystal is generated with a prescribed distribution of grain sizes, crystallographic orientations, and initial dislocation densities. A comprehensive discussion on the generation of microstructures can be found in Sec. 3.2.

During deformation under a given applied macroscopic monotonic stress, the deformation is analyzed by employing bisection methods to numerically solve the overall strain along with its upper and lower bounds. This task is achieved using a set of implicit equilibrium equations that account for aspects such as microstructure variations, dislocation density distributions per grain, strain per grain, and applied stress. This coupled system of equations enables the evaluation of the full stress–strain response of the polycrystal. Details are given in Sec. 3.3.

In this process, the predicted distributions of the dislocation density per grain are propagated through the GS-DTS equation to determine the corresponding relevant stress distributions per grain. From each stress distribution, the lower bound, mean and upper bound stresses are extracted. Details are given in Sec. 3.6.

## 3.1 Data Assembly and Preprocessing

A sufficiently large and representative data set is critical for training a robust data-driven model to predict the evolution of dislocation density in FCC metals. For this study, we focus on quasi-static tensile experiments on samples having equiaxed grains larger than 1 μm, which do not exhibit a strong texture. In addition, we focus on metals and alloys in which dislocation-mediated plasticity is observed to be the dominant deformation mechanism up to moderate strain levels.

From the literature, we collected 30 stress-strain curves for Ni [25, 42–47], 12 for Al [48–55], and 29 for Cu [56–63]. To extend the feature space to smaller grain sizes ($\leq 10$ μm) and higher strength regimes, we also included stainless steel data, in particular 3 curves for 304L, 9 for 304, 4 for 316, and 7 for 316L, respectively [64–75]. This data set included a wide range of grain sizes ($1 \leq d \leq 300$ μm), elastic properties ($26 \leq \mu \leq 76$ GPa), and yield strengths (50 MPa $\leq \sigma_y \leq 1$ GPa) as summarized in Supplementary Figure S9(a)-(c) and Supplementary Materials Section S3. A summary of the relevant material properties of these metals is summarized in Supplementary Table S1.

Each stress-strain curve was digitized every $\sim 0.01$ strain increment. Furthermore, the features reported most consistently in all experimental studies, which are generally considered to strongly influence the stress-strain curve, were recorded in the data set. These include the average grain size [76–80], strain rate [81, 82], Young's modulus, and the yield strength [24]. However, other microstructural features, such as grain size and grain orientation distributions, are rarely documented in detail, despite their known variability between experiments. As such, these features are treated here as part of the overall uncertainty. Furthermore, the average dislocation density as a function of strain is rarely reported in these experiments. Therefore, they are estimated here for each stress-strain data point using the GS-DTS model (Eq. (1)). We also estimated the dislocation density at the yield points $\rho_0$. The estimated dislocation densities versus strain are then recorded in the data set. Together, 94 experimental stress-strain curves were assembled and discretized into 3098 data points that span a wide range of grain sizes and yield strengths, as summarized in Supplementary Figure S9. Additional details on data pre-processing and data normalization are provided in Section 3.4.1. and Section 3.4.2.

## 3.2 Microstructure generation

Supplementary Figure S10 summarizes the workflow used to create synthetic polycrystals compatible with the parallel–series model. In this procedure, we begin by selecting the number of cross-sectional slices, $N_T = L/d_{\text{ave}}$, where $L$ is the total sample length in the loading direction and $d_{\text{ave}}$ is the target average grain size. Because the parallel–series framework operates on slice thicknesses rather than explicit grain sizes, we must map the desired bulk grain size distribution, $f_d(x)$, onto an equivalent thickness distribution, $f_T(T)$. Given a known cross-sectional area and the target average grain size, the total number of grains in the volume is fixed. This implies that the in-plane grain size distribution at any cross section differs from the through-thickness distribution of grain sizes in the bulk polycrystal.

To address this difference, we derive a relation between the bulk grain size distribution and the modeled thickness distribution. Specifically, as shown in the Supplementary Materials Section S2, to obtain a grain size distribution $f_d(x)$, the required probability density function for sampling slice thicknesses is:

$$f_T(T) = \frac{f_d(T)T^2}{\int_{T_{\min}}^{T_{\max}} f_d(x_d) x_d^2 \, dx_d}, \tag{3}$$

where $f_d(x)$ is the probability density function for the target grain size distribution, and $f_T(T)$ is the resulting distribution of slice thicknesses.

Using Eq. (3), we sample the thickness, $N_T$. For each slice $i$, the number of grains is then estimated as $N_i = A_0/T_i^2$, where $A_0$ is the nominal cross-sectional area. For each grain in slice $i$, we randomly generate a cross-sectional area, orientation, and initial dislocation density. Since random sampling may not exactly conserve the total cross-sectional area, we normalize the grain areas so that their sum is equal to $A_0$. Repeating this process for all $N_T$ slices yields the geometric parameters $t_i$ and $a_{ik}$ required for Eqs. (10) and (6).

Microstructural features, including grain orientations and initial dislocation densities, are sampled from prescribed distributions: The initial dislocation density in each grain is assumed to follow a uniform distribution $U[\rho_{\text{ave}} - \Delta\rho, \rho_{\text{ave}} + \Delta\rho]$, with $\Delta\rho = 0.5\rho_{\text{ave}}$. The grain orientations, expressed in terms of Euler angles $[\gamma_1, \gamma_2, \gamma_3]$, are sampled independently of a uniform distribution $U[0, 2\pi]$. Finally, grain sizes follow a log-normal distribution centered on the target average grain size $d_{\text{ave}}$.

### 3.3 Parallel-series model for polycrystals stress-strain prediction

We extend the parallel-series from "weakest-link" framework introduced by Gu et al.[10], by incorporating trained MDN to predict the strain-dependent distribution of dislocation density for every grain. Thus, this revised framework provides a data-driven micromechanics approach within a statistical homogenization scheme to predict the full polycrystalline stress–strain curve.

For the predictions of the stress-strain curve, a synthetic rectangular polycrystal is first generated having a cross-sectional area $A_0$ and height $L_0$. The microstructure characteristics (grain size distribution, crystallographic orientation, and initial dislocation density for each grain) are then extracted from the prescribed statistical distributions (Section 3.2).

The polycrystal is discretized into $N_T$ slices along the loading axis (longitudinal direction). Each slice $i \in [1, N_T]$ has thickness $T_i$ and $N_{ik}$ grains, where $(k = 1, \ldots, N_i)$. Each grain has dimensions in the plane $d_{ik}$ and a height equal to the thickness of the slice. The detailed microstructure generation is described in Section 3.2. To compute the stress–strain curve under an applied macroscopic stress $\sigma_{\text{app}}$, it is assumed that all grains inside a slice share the same axial strain, i.e., an assumption of parallel compatibility. The deformation is analyzed slice by slice with the normal strain $\epsilon_{\text{app}}^i$ in slice $i$ determined by satisfying equilibrium.

For grain $k$ in slice $i$, the MDN returns a dislocation density distribution $\rho_{ik}(\boldsymbol{\theta}_{ik}, \epsilon, \dot{\epsilon})$. The axial stress is then computed using Eq (1):

$$\sigma_{ik} = \frac{1}{M_{ik}} \left( \alpha \mu b \sqrt{\rho_{ik}} + \frac{\beta \mu}{d_{ik} \sqrt{\rho_{ik}}} \right), \tag{4}$$

where $M_{ik}$ is the maximum Schmid factor for grain $k$.

The total axial force on slice $i$ is given by summing the stress contributions over the cross-sectional area of all grains:

$$F^i = \sum_{k=1}^{N_{ik}} \sigma_{ik} A_{ik}. \tag{5}$$

and the corresponding average stress for this slice is then:

$$\tilde{\sigma}^i = \frac{F^i}{A_0} = \sum_{k=1}^{N_{ik}} \sigma_{ik} \frac{A_{ik}}{A_0} := \sum_{k=1}^{N_{ik}} \sigma_{ik} a_{ik} = \sum_{k=1}^{N_{ik}} \left[ \frac{a_{ik}}{M_{ik}} \left( \alpha \mu b \sqrt{\rho_{ik}} + \frac{\beta \mu}{d_{ik} \sqrt{\rho_{ik}}} \right) \right], \tag{6}$$

where $a_{ik} = A_{ik}/A_0$ denotes the area fraction of grain $k$. Thus, the stress-strain relation functionally can then be expressed as:

$$\tilde{\sigma}^i = F_i(\rho_{ik}(\epsilon_{\text{app}}^i \,|\, \boldsymbol{\theta}_{ik}, \dot{\epsilon})) := \sum_{k=1}^{N_{ik}} \left[ \frac{a_{ik}}{M_{ik}} \left( \alpha \mu b \sqrt{\rho_{ik}(\boldsymbol{\theta}_{ik}, \epsilon_{\text{app}}^i, \dot{\epsilon})} + \frac{\beta \mu}{d_{ik} \sqrt{\rho_{ik}(\boldsymbol{\theta}_{ik}, \epsilon_{\text{app}}^i, \dot{\epsilon})}} \right) \right]. \tag{7}$$

Because $\tilde{\sigma}^i(\epsilon)$ is monotonic in the hardening regime, an inverse function $F_i^{-1}$ exists such that

$$\epsilon_{\text{app}}^i = F_i^{-1}(\sigma_{\text{app}}). \tag{8}$$

Since a closed-form expression for $F_i^{-1}$ does not exit, Eq. (8) is solved instead numerically for each slice using the bisection method. The full procedure is summarized in Supplementary Figure S11.

Once all slice strains are determined, the overall elongation $\Delta L$ and macroscopic axial strain $\epsilon_{\text{app}}$ are computed as follows:

$$\Delta L = \sum_{i=1}^{N_T} \Delta L_i = \sum_{i=1}^{N_T} \epsilon_{\text{app}}^i T_i, \tag{9}$$

and

$$\epsilon_{\text{app}} = \frac{\Delta L}{L_0} = \sum_{i=1}^{N_T} \epsilon_{\text{app}}^i \frac{T_i}{L_0} := \sum_{i=1}^{N_T} \epsilon_{\text{app}}^i t_i = \sum_{i=1}^{N_T} F_i^{-1}(\sigma_{\text{app}}, \rho_{ik}) t_i. \tag{10}$$

This parallel–series framework links microstructural statistics directly to macroscopic stress–strain behavior, demonstrating that the MDN-predicted dislocation density evolution can be embedded in a conventional polycrystal solver with minimal additional assumptions. In addition, although the present implementation relies on uniform-strain slices and a size-dependent Taylor hardening law, the same MDN output could be coupled with more sophisticated homogenization schemes (e.g., self-consistent or FFT-based models) without retraining.

### 3.4 Data-driven Model for dislocation density evolution

#### 3.4.1 Data collection and preprocessing

We consider the five properties most frequently reported in the literature as features of our model, including Young's modulus $E$, mechanical response (strain $\epsilon$, strain rate $\dot{\epsilon}$ and yield stress $\sigma_y$), and microstructures (grain size $d$).

Supplementary Figure S13 shows a schematic overview of the data preprocessing workflow used to generate the training data set from stress-strain curves reported in the literature. Each curve is digitized into discrete stress and strain values using a strain increment of approximately 0.01. The yield stress is extracted directly from the source text or estimated from the curve at the point where the slope of a stress-strain curve noticeably decreases. Additional information such as strain rate, grain size, and elastic modulus is extracted from the accompanying text or tables. All collected features are then consolidated into a structured data set, which forms the input for the model training. The target output of the model is the dislocation density, calculated using Eq. (1). It should be noted that this equation is not strictly monotonic, meaning that two different dislocation density values can correspond to the same stress level. In such cases, we select the solution with the highest dislocation density. This choice is justified by the absence of significant softening behavior and the average grain size being larger than 1 μm in the experimental data [24]. If the lower solution were correct, continued loading would result in an increase in dislocation density and a decrease in stress, indicating softening, contrary to what is observed in the current experimental set. Since this study focuses on strain hardening mechanisms, we adopt the higher-value solution throughout.

#### 3.4.2 Data normalization

Supplementary Figure S9 shows that the raw input data span several orders of magnitude and do not follow any specific distribution. To manage this variability, we applied a quantile transformation to convert the distributions of strain ($\epsilon$), strain rate ($\dot{\epsilon}$), and the average grain size ($d_{ave}$) into standard uniform distributions. The Young's modulus ($E$) and yield strength ($\sigma_y$) are normalized using a Min-Max Scaler.

**Table 1** Hyperparameter tuning for the MDN model.

| Hyperparameter | Learning Rate ($l_r$) | Number of Layers | Neurons per Layer | Number of Gaussian ($K$) |
|---|---|---|---|---|
| Interval | [1e-5, 0.1] | [1, 10] | [1, 10] | [1, 10] |
| Optimal Value | 0.00145 | 10 | 4 | 2 |

For the output of the model, the dislocation density distribution, which ranges from approximately $10^{12}\,\text{m}^{-2}$ to $10^{14}\,\text{m}^{-2}$, we first apply a logarithmic transformation to reduce skewness, followed by Min-Max scaling to map values to the interval $[0,1]$. These fitted scalers are retained and later reused to normalize new input data during inference.

### 3.5 Training of the mixed density network

Given an input feature vector $\mathbf{x} = \{\epsilon, \dot{\epsilon}, E, d, \sigma_y\}$, we model the conditional probability distribution of the dislocation density $\rho$ as a convex combination of $K$ elemental distributions with corresponding parameters $\theta_k(\mathbf{x})$ as follows:

$$\phi(\rho|\mathbf{x}) = \sum_{k=1}^{K} \Pi_k(\mathbf{x}) \phi_k(\rho, \theta_k(\mathbf{x})) \tag{11}$$

Here, $\Pi_k(\mathbf{x})$ are the mixing coefficients that satisfy the constraint $\sum_{k=1}^{K} \Pi_k(\mathbf{x}) = 1$. Each component $\phi_k$ is a Gaussian distribution parameterized by $\theta_k = \{M_k, \Sigma_k\}$, where $M_k$ and $\Sigma_k$ represent the mean and standard deviation of the distribution, respectively. As such, the MDN output is a mixture of $K$ Gaussian conditioned on the input $\mathbf{x}$.

The MDN is trained by minimizing the negative log-likelihood of the observed dislocation densities:

$$\mathcal{L} = -\ln\left[\prod_{i=1}^{N} \phi(\rho_{i,ave}|\mathbf{x_i})\right], \tag{12}$$

where $N$ is the total number of training samples. Key hyperparameters, such as the learning rate $l_r$, the number of layers in the network, the number of neurons per layer, and the number of Gaussian components $K$, are optimized using a grid search strategy combined with the Adam optimizer [83] and a 10-fold cross-validation to prevent overfitting. The search ranges and optimal values are summarized in Table 1.

During hyperparameter tuning, it was consistently observed that $K = 2$ provides the most robust performance. As such, this value was fixed in the final tuning stage to streamline the optimization of the remaining hyperparameters.

### 3.6 Uncertainty Quantification: Upper and Lower Stress–Strain Bounds

In Eq. (10), the dislocation density $\rho_{ik}(\boldsymbol{\theta}, \epsilon, \dot{\epsilon}, T)$ is predicted using the trained MDN model as a probability distribution. Consequently, the stress $\sigma_{ik}$ is itself a random variable with a distribution that reflects the uncertainty.

To build a numerical approximation of the probability density of stress $p_\sigma(\sigma_{ik})$, we extract samples from the MDN-predicted $\rho_{ik}$ and then propagate them through the GS-DTS to obtain samples of $\sigma_{ik}$. The confidence bounds can be determined as follows. Let $M[\cdot]$ and $\Sigma[\cdot]$ denote the mean and standard deviation of $p_\sigma(\sigma_{ik})$. The upper bound, lower bound, and average stress can then be defined as

$$\sigma_{ik}^{\text{upper}} = M[p_\sigma(\sigma_{ik})] + R\,\Sigma[p_\sigma(\sigma_{ik})], \qquad \sigma_{ik}^{\text{lower}} = M[p_\sigma(\sigma_{ik})] - R\,\Sigma[p_\sigma(\sigma_{ik})], \qquad \sigma_{ik}^{\text{avg}} = M[p_\sigma(\sigma_{ik})]. \tag{13}$$

where $R$ is the confidence multiplier. Here, we use $R = 2$ for Ni and Cu according to the Empirical Rule in Statistics, and a more conservative $R = 1.5$ for Al to account for the smaller and less consistent data available in literature. For NiCoCr and NiCoCrMnFe, a value of $R = 1.5$ is also chosen to keep the prediction conservative.

The upper, lower, and average macroscopic stresses can then be obtained by substituting $\sigma_{ik}^{\#}$ ($\# \in \{\text{upper}, \text{lower}, \text{avg}\}$) for $\sigma_{ik}$ in Eq. (6). The result is therefore a complete stress–strain envelope that quantifies the predictive uncertainty of the model at every strain level.

## 4 Results

### 4.1 Mixture Density Network for Dislocation Density Evolution

Figure 2(a) shows the normalized dislocation density predicted by the trained MDN (mean ± standard deviation) versus those estimated from the experimental stress-strain curves using the GS-DTS model. The close agreement of the mean predicted values with the experimental estimates suggests that the trained MDN is highly effective in capturing the overall distribution of dislocation densities for the wide range of microstructures and strain ranges in our data set. Furthermore, the standard deviation bars reflect the inherent scatter in the experimental results due to uncertainty for the same input feature vector $\mathbf{x}$. However, it should be noted that the MDN mean predictions deviate from those estimated from the experiments for $\rho < 1.78 \times 10^{12}$ m$^{-2}$. This is due to the small number of experimental results in the training data set for this low dislocation density regime as compared to the higher dislocation density regime.

Figures 2(b), (c), and (d) show comparisons between MDN predictions of the dislocation density distributions for Cu, Al, and Ni for different grain sizes at $\epsilon = 0.1$ and $\rho_0 = 10^{13}$ m$^{-2}$. Each distribution captures the inherent experimental variability due to uncertainty and possible inconsistencies in the microstructural features that are not reported experimentally. The results also indicate that both Ni and Cu exhibit a bimodal distribution at intermediate strain levels, while those for Al show

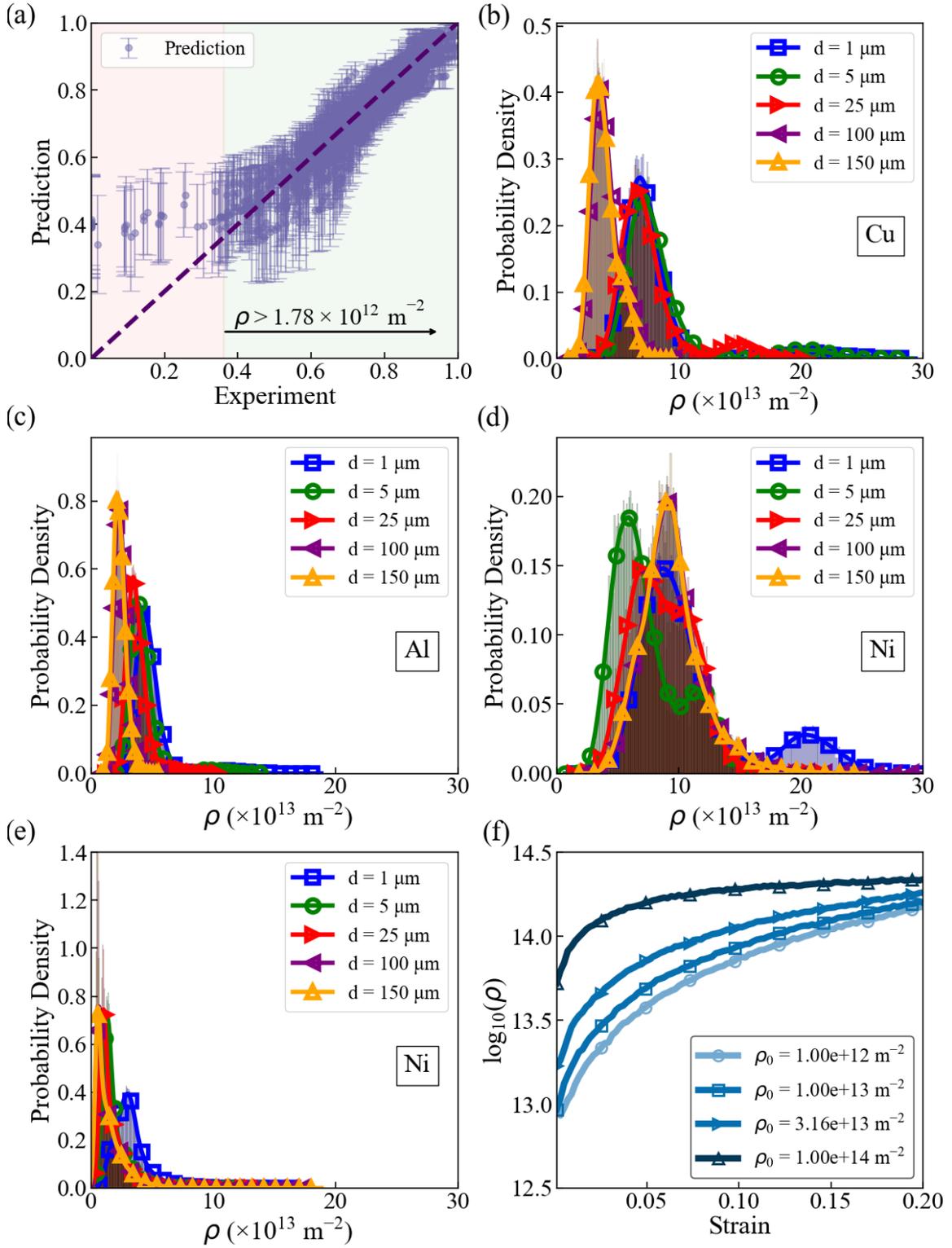

**Fig. 2 MDN Predictions for Dislocation Density.** (a) The MDN predicted mean $\pm$ standard deviation of the normalized dislocation densities versus experimental values derived using Eq. (1) for the entire dataset. (b)-(d) The MDN predicted dislocation density distributions for Cu, Al, and Ni, respectively, having different grain sizes, $\rho_0 = 10^{13}$ m$^{-2}$, and at $\epsilon = 0.1$. (e) The MDN predicted dislocation density distributions for Ni having different grain sizes, $\rho_0 = 10^{13}$ m$^{-2}$, and at $\epsilon = 0.01$. (f) The MDN predicted mean values for the evolution of the dislocation density in Ni having grain size $d = 25\mu$m versus strain for different initial dislocation densities.

a normal distribution. The difference could be attributed to the smaller number of Al curves in the data set and intrinsically reflect the combined effects of aleatory and epistemic uncertainties in the experimental data, which arise from differences in microstructural features, measurement techniques, and unreported experimental conditions. These distinctive patterns highlight the importance of accounting for these distributions rather than relying on a single deterministic value to adequately capture the uncertainties.

Figures 2(d) and (e) also show the comparison between MDN predictions of the dislocation density distributions at $\epsilon = 0.1$ and $\epsilon = 0.01$, respectively, for different grain sizes of Ni and an initial dislocation density of $\rho_0 = 10^{13}$ m$^{-2}$. These clearly show that the distributions broaden and shift as a function of both grain size and strain. In particular, this variation points to the increase in uncertainty in experiments with increasing strain.

These findings illustrate the substantial uncertainties in existing experimental data sets in the literature and reveal the need for more rigorous characterization of the underlying microstructure to enable the development of more deterministic and physically grounded models for the evolution of dislocation density in the future.

Figure 2(f) further shows the effect of the initial dislocation density on the evolution of the mean value of the dislocation density as a function of strain. The results are shown for Ni with a grain size of 25 μm. It is observed that lower initial dislocation densities lead to higher rate of multiplication of the dislocation density at low strains. Ultimately, at higher strain levels, all curves reach the same asymptotic behavior irrespective of the initial dislocation density [31].

Finally, Figs. 3(a) and (b) show a comparison between the MDN-predicted evolution of the dislocation density and the experimentally measured values from [25] for $d = 32$ μm and $d = 90$ μm, respectively. The solid lines represent the predicted mean dislocation density, while the shaded regions indicate the associated uncertainty bounds. The experimental data points for both grain sizes clearly fall within the predicted uncertainty range throughout the strain range, demonstrating that the MDN model effectively captures both the trend and the variability of the evolution of dislocations across different grain sizes, in excellent agreement with experiments. This agreement highlights the model's ability to generalize and accommodate experimental uncertainties without retraining or recalibration.

### 4.2 MDN-Based Prediction of Polycrystalline Stress–Strain Behavior

By combining the MDN predicted dislocation density distributions per grain with the GS-DTS law (Eq. 1), we can calculate the corresponding single-grain shear stress distribution at any strain level, as shown in Fig. 4(a) for a Ni grain with $d = 25$ μm and $\rho_0 = 10^{13}$ m$^{-2}$ at four different strain levels.

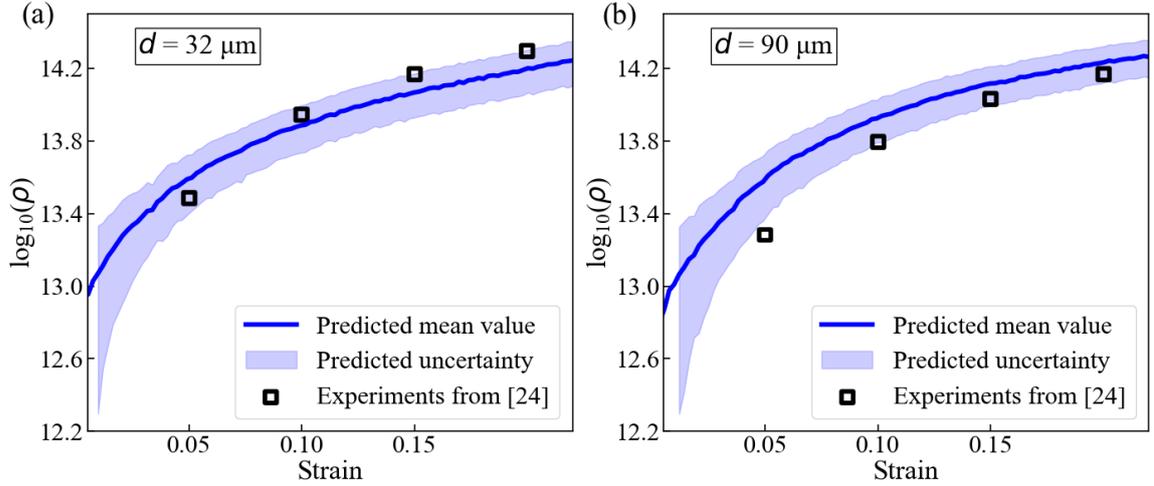

**Fig. 3** Comparison between the MDN-predicted dislocation density and experimentally measured values from [25] for polycrystalline Ni having: (a) average grain size $d = 32$ μm; and (b) average grain size $d = 90$ μm.

These stress distributions show patterns similar to the dislocation density distributions discussed earlier and represent the uncertainty in our predictions.

To simulate polycrystalline stress-strain curves, we employ a parallel-series model [10] in which the sample is subdivided along the tensile axis into multiple "slices", each containing many grains with stochastically assigned grain sizes, orientations, and initial dislocation densities as described in Section 3.2. This model allows different slices to develop different local strains, while ensuring that the overall force is balanced. The overall stress-strain curve is then predicted from the force balance in the slices and the local strain in each slice (see Methods, Section 3.3).

Figure 4(b)-(d) show representative predicted stress–strain curves for polycrystalline Ni, Cu, and Al, respectively. Each plot includes the predicted mean response along with the upper and lower uncertainty bounds, compared to the corresponding experimental data. In figure 4(b), the simulated Ni polycrystal has an average grain size $d_{\text{avg}} = 2$ μm and an average initial dislocation density $\rho_{0,\text{avg}} = 4.83 \times 10^{13}$ m$^{-2}$ estimated from experiments using the GS-DTS model. Furthermore, the synthetic polycrystal was constructed from a total of 45692 grains in total, ensuring a large representative volume of the experiment. In Figure 4(c), the Cu polycrystal has $d_{\text{avg}} = 14.5$ μm, $\rho_{0,\text{avg}} = 2.02 \times 10^{13}$ m$^{-2}$ estimated from the experiment, and 46390 grains. In Figure 4(d), the Al polycrystal has $d_{\text{avg}} = 12.0$ μm, $\rho_{0,\text{avg}} = 9.39 \times 10^{12}$ m$^{-2}$, and 46096 total grains.

In all calculations, a log-normal grain size distribution with the standard deviation of the grain size distribution being $0.01 \times d_{ave}$, a uniform initial dislocation density distribution with standard deviation $\rho_{0,ave}/2\sqrt{3}$, a uniform grain orientation distribution of $[0, 2\pi] \times [0, 2\pi] \times [0, 2\pi]$ for its Euler angles, and a Taylor factor of 3.06, were used. Additional predictions for the entire pure metal data set are provided in Supplementary Figures S4 through S6 for completeness.

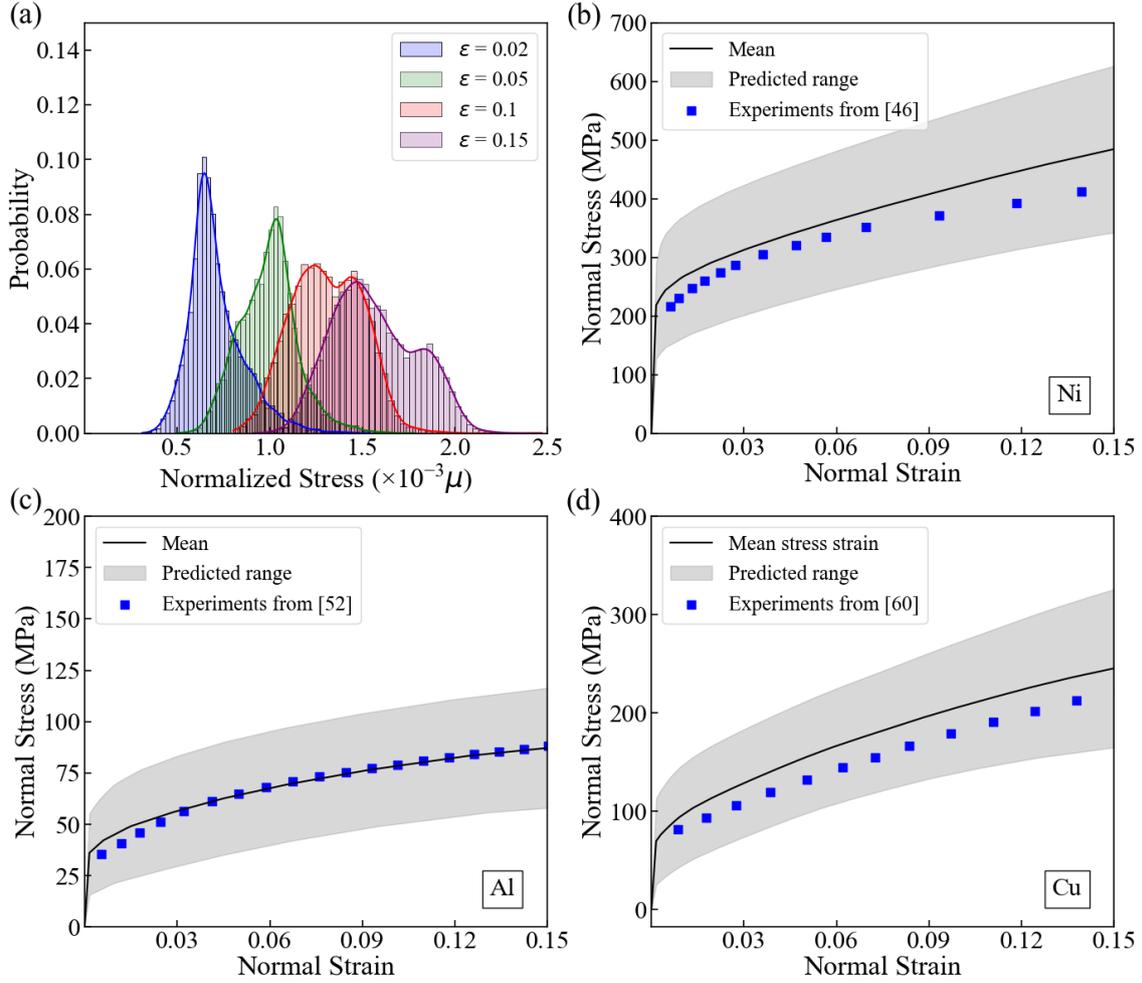

**Fig. 4 From the predictions of single-grain stress distributions to polycrystalline stress-strain curve predictions with uncertainty.** (a) Shear stress distributions at $\epsilon = 0.02, 0.05, 0.1$, and $0.15$ for a single Ni grain having $d = 25$ μm and $\rho_0 = 10^{13}$ m$^{-2}$. (b) Predicted stress-strain curve (mean value and uncertainty range) for a Ni polycrystal having $d_{\text{avg}} = 2$ μm and $\rho_{0,\text{avg}} = 4.83 \times 10^{13}$ m$^{-2}$. (c) Predicted stress-strain curve (mean value and uncertainty range) for a Cu polycrystal having $d_{\text{avg}} = 14.5$ μm and $\rho_{0,\text{avg}} = 2.02 \times 10^{13}$ m$^{-2}$. (d) Predicted stress-strain curve (mean value and uncertainty range) for an Al polycrystal having $d_{\text{avg}} = 12.0$ μm and $\rho_{0,\text{avg}} = 9.32 \times 10^{12}$ m$^{-2}$.

To quantitatively assess the agreement between our model and the experimental data in the entire data set, we define a Mahalanobis-like distance measure $\mathscr{D}$, which compares the experimental data $(\hat{\sigma}_i, \hat{\epsilon}_i)$ with the model mean and variance predictions as follows:

$$\mathscr{D} = \sqrt{\frac{1}{n} \sum_{i=1}^{n} \frac{\left(\sigma_{i,ave} - \hat{\sigma}_i\right)^2}{V_i^2}}, \qquad (14)$$

where $\sigma_{i,ave}$ is the predicted mean stress at each strain increment $\hat{\epsilon}_i$, $V_i = \frac{1}{2}\left(\sigma_{i,ave}^{\text{upper}} - \sigma_{i,ave}^{\text{lower}}\right)$ is half the width of the predicted confidence interval at $\hat{\epsilon}_i$, $n$ is the number of data points used to discretize the experimental stress-strain curve, and $\hat{\sigma}_i$ is the experimentally measured stress at $\hat{\epsilon}_i$. It should be noted that $\mathscr{D} = 0$ indicates that the mean value matches perfectly with the experimental data, while $\mathscr{D} > 1$ indicates that the experimental measurement lies outside of our predicted uncertainty range.

Figures 5(a) and (b) show that most of the stress-strain curves predicted for pure metals have $\mathscr{D} < 1$, indicating that the experiential curves are well within the predicted uncertainty range. Only one Ni, one Cu, and one Al stress-strain curves from the data set of a total of 64 pure metal stress-strain curves were outside the range of our predictions. There is also no systematic dependency between $\mathscr{D}$ and grain size, as shown in Figure 5(b). This suggests that the current model captures microstructural variability without systematic bias.

These results show that without any parameter adjustment or case-specific calibration, the model accurately reproduces the stress–strain behavior of three chemically and mechanically distinct FCC metals in a wide range of grain sizes. In each case, the experimental curves fall within the predicted uncertainty range.

Having established that our model accurately predicts stress–strain behavior for pure FCC metals within the training dataset, we now turn to a critical question: Under what conditions does this model remain valid when applied to more complex alloy systems? Because the model is developed to capture the fundamental mechanisms governing dislocation evolution in FCC metals, we hypothesize that it should generalize to other FCC alloys, provided their deformation behavior remains primarily controlled by dislocation-mediated plasticity. In other words, the model should remain accurate for any FCC material in which dislocation plasticity is the dominant factor in the mechanical response.

To test this hypothesis, the trained dislocation density evolution model is applied without any additional training or calibration to predict the stress-strain curves of two multicomponent FCC alloys: NiCoCr and NiCoCrMnFe. These alloys exhibit additional solute strengthening, which can be accounted for by computing the appropriate value of $\tau_0$ for each alloy according to Eq. (6). These values were estimated using the solute strengthening model developed by Leyson et al. (2012) [84] and calculated in [85]. The estimated values for $\tau_0$ for both alloys are summarized in Supplementary Table S1. To validate the predictions of the model, we have also conducted tensile experiments on both alloys.

Figure 5(c) and (d) show two representative stress–strain curves predicted by the model alongside the corresponding experimental data. Additional results for both alloys at different average grain sizes are provided in Supplementary Figures S7 and S8 for completeness. As shown in Fig. 5(e) and (f), all $\mathscr{D}$ values for the eight NiCoCr and ten NiCoCrMnFe examined here are smaller than 1. In addition, the relationship between the distribution of $\mathscr{D}$ and both the initial dislocation density and the yield stress is not statistically significant, similar to what we have observed in pure FCC metals.

This result highlights the strong generalizability of the model, confirming its ability to accurately predict the mechanical response of more complex FCC alloys without requiring additional calibration or parameter fitting. These findings provide compelling evidence for the robustness of the framework

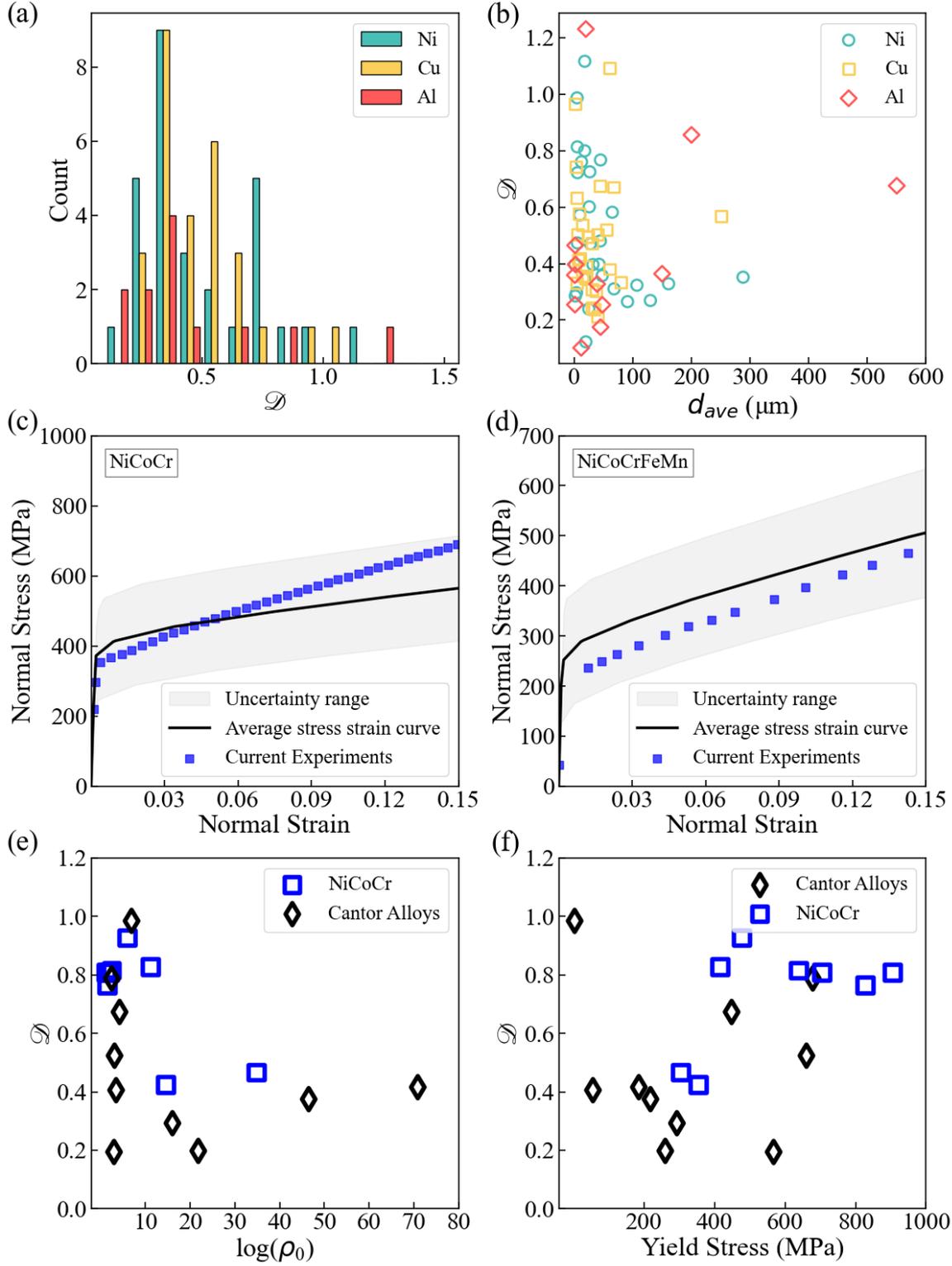

**Fig. 5** (a) Distribution of $\mathscr{D}$ for the pure metals in the data base. (b) $\mathscr{D}$ versus average grain size for the pure metals dataset. (c) A predicted stress strain curve and its uncertainty range for a Cantor alloy having $d_{\text{ave}} = 70.90$ μm and an initial dislocation density $\rho_0 = 2.61 \times 10^{12} \text{m}^{-2}$. The corresponding stress-strain curve measured experimentally in this study is also shown for comparison. (d) The predicted stress strain curve and its uncertainty range for a NiCoCr alloy having $d_{\text{ave}} = 14.56$ μm and an initial dislocation density of $\rho_0 = 3.27 \times 10^{13} \text{m}^{-2}$. The corresponding stress-strain curve measured experimentally in this study is also shown for comparison. (e) $\mathscr{D}$ versus average grain size for NiCoCr and NiCoCrMnFe alloys. (f) $\mathscr{D}$ versus the yield stress for NiCoCr and NiCoCrMnFe alloys.

and indicate that the underlying statistical treatment of dislocation evolution and microstructural variability effectively captures the essential physics across a wide range of FCC material systems, in which dislocation plasticity is the dominant deformation mechanism.

## 5 Conclusions

The results presented in this study demonstrate that our data-driven framework, built on a Mixed Density Network (MDN) trained with data based on the literature, effectively captures the evolution of dislocation density and predicts the stress–strain response of a wide range of FCC metals. Remarkably, the model achieves high fidelity across diverse compositions and microstructures without any per-case parameter fitting or recalibration. This includes successful application to chemically complex multicomponent systems such as NiCoCr and NiCoCrMnFe.

These findings emphasize the importance of incorporating uncertainty and statistical variability into microstructure-based models. Traditional deterministic frameworks often fail to reconcile the scatter observed in experimental stress–strain data, especially when critical microstructural features, such as grain orientation distributions or local dislocation densities, are not reported. Through inference of complete probabilistic distributions rather than single-value calculation, our model accounts for both aleatory and epistemic uncertainties, thereby enabling robust and reliable predictions despite incomplete microstructural information.

It is also demonstrated that the model can be generalized to new FCC alloys, provided that dislocation-mediated plasticity remains the primary deformation mechanism, highlighting the importance of mechanism-aware learning in developing transferable predictive frameworks. Furthermore, the successful predictions made by this model suggest that the framework can serve as a reliable backbone for high-throughput alloy screening and property prediction in compositionally complex materials.

Despite its strengths, the model's predictive uncertainty increases at higher strain levels or low initial dislocation density, reflecting the limited availability of experimental data from those regimes in the training set. Incorporating additional datasets, particularly those that capture large plastic deformations and low initial dislocation density, as well as more comprehensive characterization of the three-dimensional microstructure, would greatly improve the model's fidelity in this regime.

Beyond its predictive capabilities, the model also provides valuable guidance for experimental protocols. By explicitly quantifying uncertainty and identifying which microstructural features contribute the most to variability in the stress–strain response, the framework can inform experimentalists on what measurements are most critical to report, such as initial dislocation density, grain size distribution, or orientation spread. In addition, when experimental results fall significantly outside the predicted uncertainty limits of the model, they may indicate potential measurement errors, inconsistencies in sample preparation, or unreported features such as texture, residual stresses, or secondary phases. In this way, the model serves not only as a predictive engine but also as a feedback tool for improving experimental reproducibility and microstructural characterization.

Overall, this work underscores the strong potential of uncertainty-aware, physics-informed machine learning models to bridge experimental limitations and enable robust, predictive simulations for new alloy design.

# Acknowledgment

Research was sponsored by the Army Research Laboratory and was accomplished under Cooperative Agreement Number W911NF-23-2-0062. The views and conclusions contained in this document are those of the authors and should not be interpreted as representing the official policies, either expressed or implied, of the Army Research Laboratory or the U.S. Government. The U.S. Government is authorized to reproduce and distribute reprints for Government purposes notwithstanding any copyright notation herein.

# Declarations

# Supplemental Materials for "Uncertainty-Aware Machine-Learning Framework for Predicting Dislocation Plasticity and Stress–Strain Response in FCC Alloys"

Jing Luo[1], Yejun Gu[2,1], Yanfei Wang[3], Xiaolong Ma[4], Jaafar A. El-Awady[1*]

[1*]Department of Mechanical Engineering, Johns Hopkins University, 3400 N Charles Street, Baltimore, 21218, MD, USA.
[2]Institute of High Performance Computing, Agency for Science, Technology and Research, 138632, Singapore.
[3]State Key Laboratory of Subtropical Building Science & Department of Engineering Mechanics, South China University of Technology, Guangzhou,510640, China.
[4]Department of Materials Science and Engineering, City University of Hong Kong, Hong Kong, China.

*Corresponding author(s). E-mail(s): jelawady1@jhu.edu;

## S1 The generalizability of the KME model

Fig. S1(a) illustrates the variability in the experimentally measured stress-strain curves for pure Ni with an average grain size of $d_{\text{ave}} = 18$ μm [S1, S2, S3], Cu with $d_{\text{ave}} = 40$ μm [S4, S5], and Al with $d_{\text{ave}} = 20$ μm [S6, S7].

Furthermore, the classical Taylor strengthening law can be expressed as

$$\tau' = \tau_0 + \alpha\mu b\sqrt{(\rho)} \tag{S1}$$

where $\tau'$ is the resolved shear stress, $\tau_0$ is critical shear stress. $\alpha$ is a material parameter, $\mu$ and $b$ are shear modulus and magnitude of Burger's vector respectively.

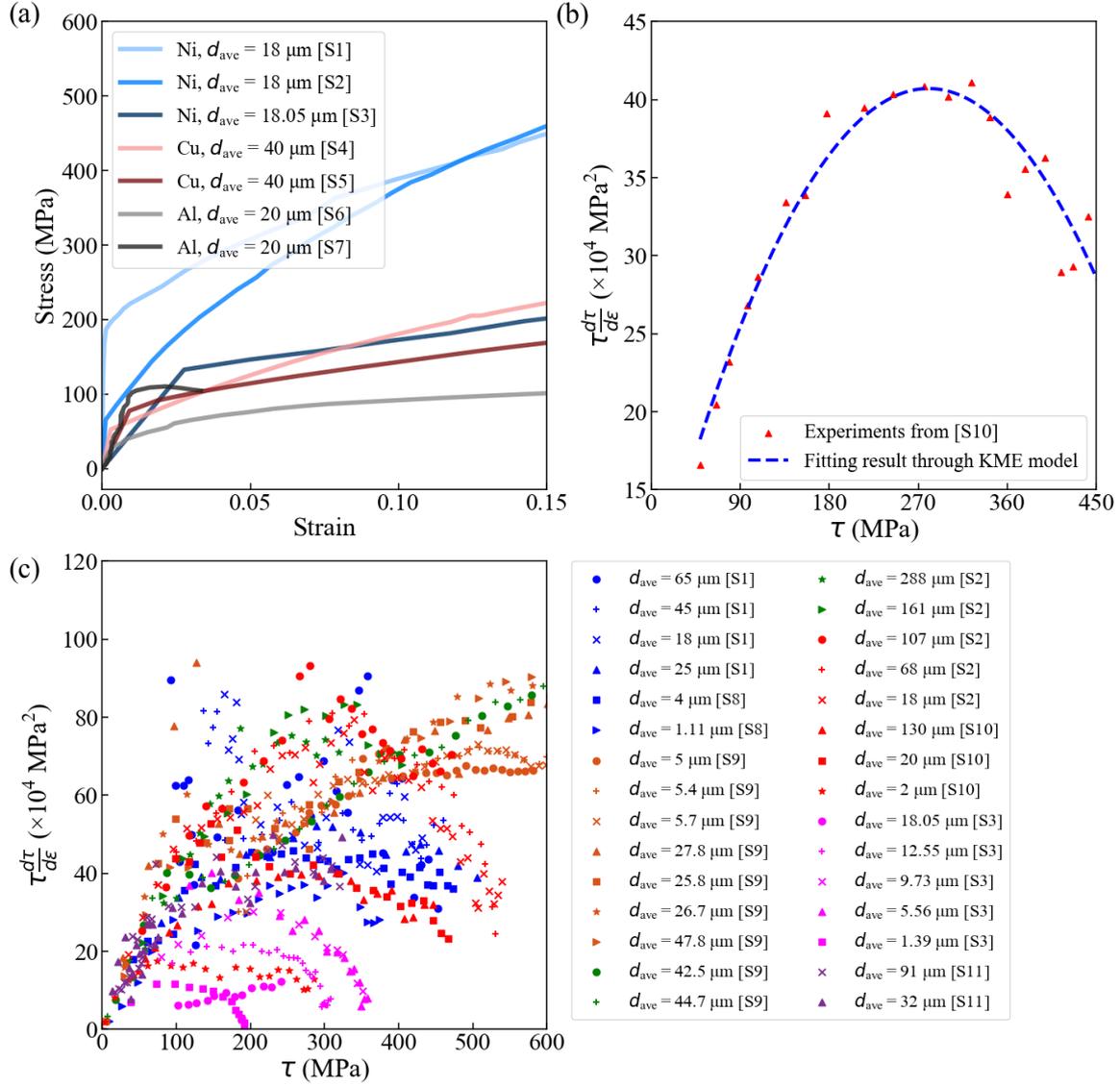

**Fig. S1** Uncertainty in experimental results and the KME model. (a) Experimental stress strains curves for pure Ni with $d_{\text{ave}} \sim 18$ μm, pure Cu with $d_{\text{ave}} \sim 40$ μm, and pure Al with grain size $d_{\text{ave}} \sim 20$μm showing the significant variability from one study to another for the same material and the same average grain size. (b) Quadratic function fitting for the KME model parameters according to Eq. (S2) from a single experimental stress-strain curve for pure Ni with $d_{\text{ave}} = 130$ μm. (c) Experimental data for pure Ni at different initial dislocation densities and grain sizes from multi sources showing that no single fitting for the KME model parameters exist.

The fitting parameters $K_0$, $K_1$, and $K_2$ in the KME model can be predicting from a given experimental stress strain curve by combining Eq. (S1) with Eq. (2) in the manuscript, which gives

$$\tau \frac{\mathrm{d}\tau}{\mathrm{d}\epsilon} = \frac{1}{2}\left(K_0(\alpha\mu b)^2 d^{-1} + K_1 \frac{\tau}{\alpha\mu b} - K_2 \tau^2\right), \tag{S2}$$

where $\tau = \tau' - \tau_0$.

From Eq. (S2), $\tau \mathrm{d}\tau/\mathrm{d}\epsilon$ should be a quadratic function of $\tau$ if the KME model correctly describes the physics of the evolution of the dislocation density.

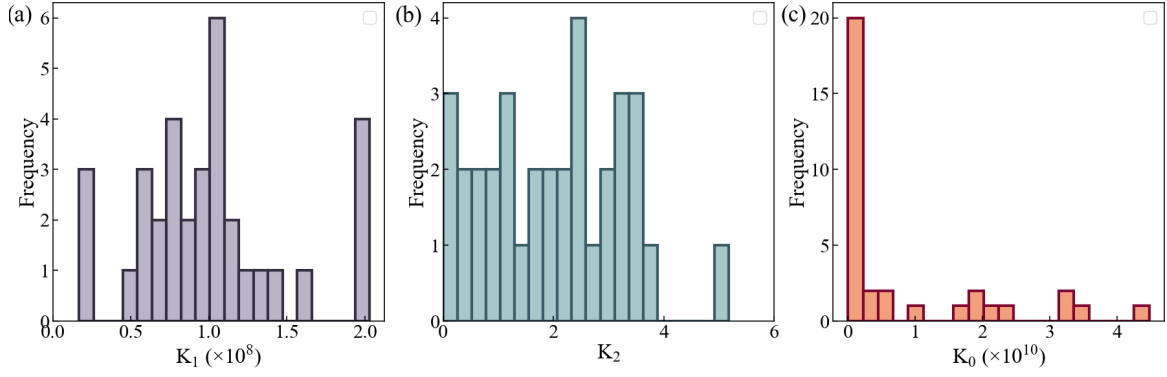

**Fig. S2** Histograms of the fitting parameters for (a) $K_0$, (b) $K_1$ and (c) $K_2$.

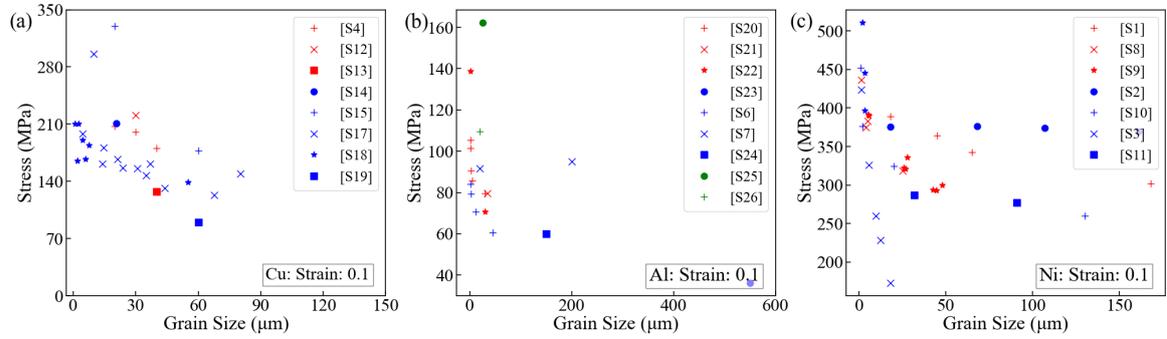

**Fig. S3** Hall-Patch relationships for (a) Cu [S4, S12, S13, S14, S15, S16, S17, S18, S19], (b) Al [S20, S21, S22, S23, S24, S6, S7, S25, S26], and (c) Ni [S1, S8, S9, S2, S10, S3, S11] at strain level 0.1.

Fig. S1(b) shows a comparison between the fitting of the KME model parameters based on a single pure Ni experiment with $d_{ave} = 18$ μm, according to the quadratic form in Eq. (S2) [S2]. The comparison indicates that by choosing the appropriate fitting parameters, the KME model can closely replicate the results of an individual experiment. However, when experimental data from multiple sources are used, even for the same material, there is no single set of fitting parameters that match all experiments. As shown in Fig. S1(c), the experimental data for Ni exhibit considerable scatters [S1, S8, S9, S2, S10, S3, S11]. The set of fitting parameters $K_0$, $K_1$, and $K_2$ for the KME model performed for each experiment one at a time are summarized in Figs. S2(a)-(c). It is clear that each parameter spans a wide range that varies from one experiment to another. This suggests that the KME model does not naturally capture all possible microstructural variations and is unsuitable for designing new materials because we cannot pre-determined its fitting parameters in a narrow range.

In Figs. S3(a)-(c), inconsistencies among experimental data are also observed to be prevalent throughout the literature in the Hall-Petch relationship for Cu, Al, and Ni.

## S2 The slice thickness probability density distribution

If the cross-sectional area of a sample is $A_0$ and the thickness of the $k$-th slice is $d_k$, the distribution of grain sizes in this section should have a mean value of $d_k$, assuming that the grains are of equiaxed shapes. Therefore,

$$A_0 = E\left[C_0 \sum_{i=1}^{N_k} (d_i^2)\right] \tag{S3}$$

where $C_0$ is a shape constant of order $O(1)$. Setting $C_0 = 1$, will approximate the grains as cubes. In addition, it is assumed that the grains in this section are independent and identically distributed. Thus,

$$A_0 = E\left[\sum_{i=1}^{N_k} (d_i^2)\right] = \sum_{i=1}^{N_k} E\left[(d_i^2)\right] = N_k(d_k^2 + \Sigma_k^2) \tag{S4}$$

where $C_0$ was set to 1 and $\Sigma_k$ represents the standard deviation of the grain size distribution in the $k$-th slice. If one choose $\Sigma_k = C_1 d_k$, then

$$N_k = \frac{1}{1+C_1^2} \frac{A_0}{d_k^2} := C_2 \frac{A_0}{d_k^2}. \tag{S5}$$

This indicates that in each slice there are $N_k$ grains, each with an expected grain size of $d_k$. Assuming the variation in grain size within a single slice is less pronounced than the variation in slide height, the probability $p_d(x_d)$ of obtaining a grain size $x_d$ across all $N$ slices in the bulk sample is related to the probability $p_T(x_d)$ of obtaining a thickness $x_d$ through

$$p_d(x_d) = \frac{C_2 \frac{A_0}{x_d^2} p_T(x_d)}{\sum_{k=1}^{N}\left(C_2 \frac{A_0 p_T(x_k)}{x_k^2}\right)} = \frac{\frac{p_T(x_d)}{x_d^2}}{\sum_{k=1}^{N} \frac{p_T(x_k)}{x_k^2}}. \tag{S6}$$

The continuous version of this distribution is then expressed as

$$f_d(x_d) = \frac{f_T(x_d)}{x_d^2} \frac{1}{\int_{x_{min}}^{x_{max}} \frac{p_T(x_d)}{x^2} dx}. \tag{S7}$$

Given that $\int_{x_{min}}^{x_{max}} \frac{p_T(x_d)}{x^2} dx$ is a constant for normalization, the equation can be written as:

$$f_T(x_d) = C f_d(x_d) x_d^2. \tag{S8}$$

Notice, $f_T$ is a probability density function and its integral over its domain should equal 1, leading to

$$C = \frac{1}{\int f_d(x_d) x_d^2 dx_d}. \tag{S9}$$

Through Eq.(S8) and Eq. (S9), we can convert the distribution of grain size in the bulk $f_d(x_d)$ with th e distribution of thickness for slices $f_T(x_d)$.

## S3 Comparisons of the predicted stress strain curves and the corresponding experiment for pure metals collected from literature as well as those for the NiCoCrMnFe and NiCoCr conducted in this study

Table S1  Material properties for the FCC metals analyzed in this study.

| Material | Shear modulus (GPa) | Poisson's ratio | Lattice constant (nm) | CRSS (MPa) |
|---|---|---|---|---|
| Al | 26 | 0.35 | 0.4046 | 0 |
| Cu | 48 | 0.34 | 0.3596 | 0 |
| Ni | 76 | 0.31 | 0.3499 | 0 |
| 316 & 316L [S27] | 84 | 0.305 | 0.37 | 24 |
| 304 & 304L [S27] | 78 | 0.305 | 0.37 | 22 |
| Cantor alloy [S28] | 80 | 0.265 | 3.59 | 40.8 |
| NiCoCr [S28] | 87 | 0.3 | 3.559 | 45.0 |

Table S2  Hyperparameter tuning for the MDN model.

| Hyperparameter | Learning Rate ($l_r$) | Number of Layers | Neurons per Layer | Number of Gaussian ($K$) |
|---|---|---|---|---|
| Interval | [1e-5, 0.1] | [1, 10] | [1, 10] | [1, 10] |
| Optimal Value | 0.00145 | 10 | 4 | 2 |

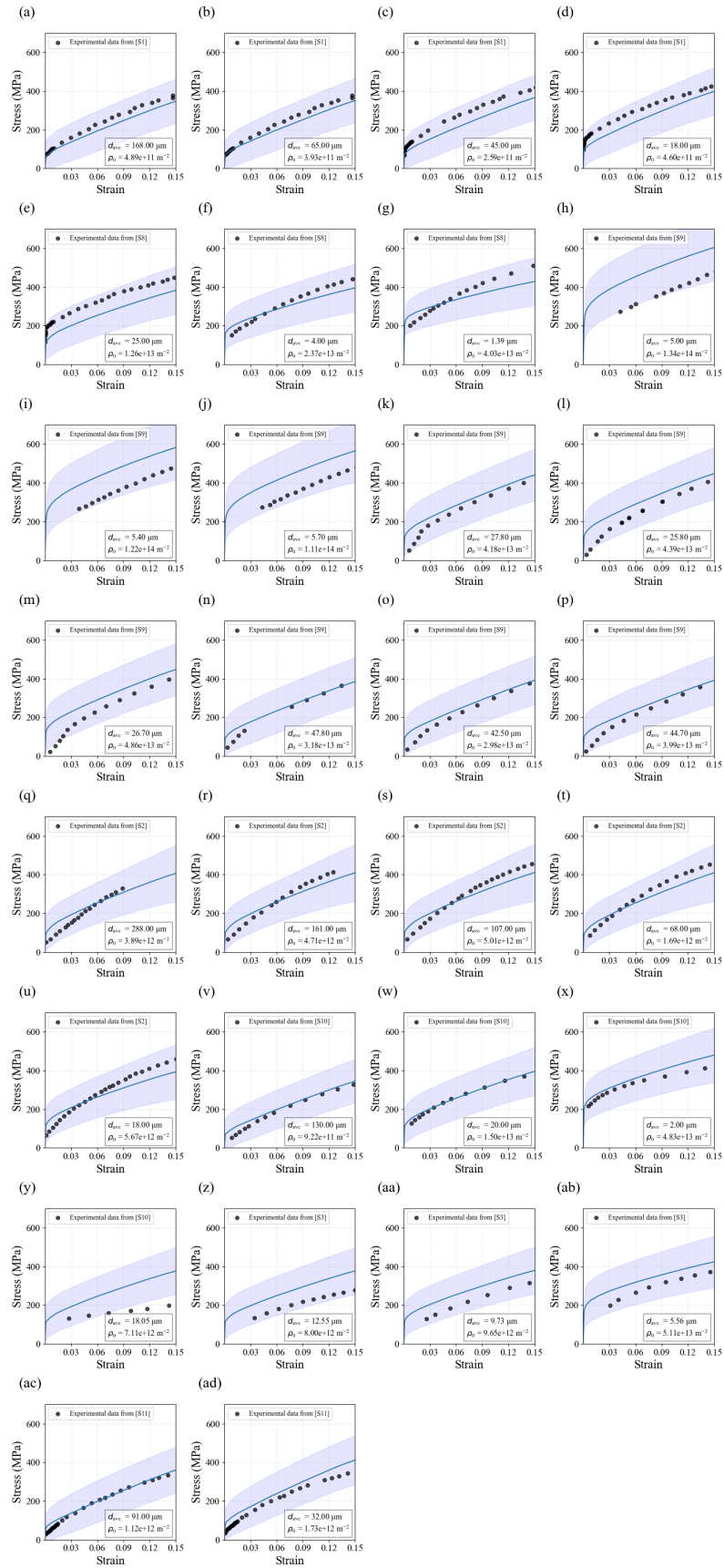

**Fig. S4** Comparisons of the predicted stress strain curves and the corresponding experiment for pure Ni collected from literature. In all figures the mean is shown by the solid line and the uncertainty range is shown by the shaded region. The corresponding experimental result is shown as solid circles.

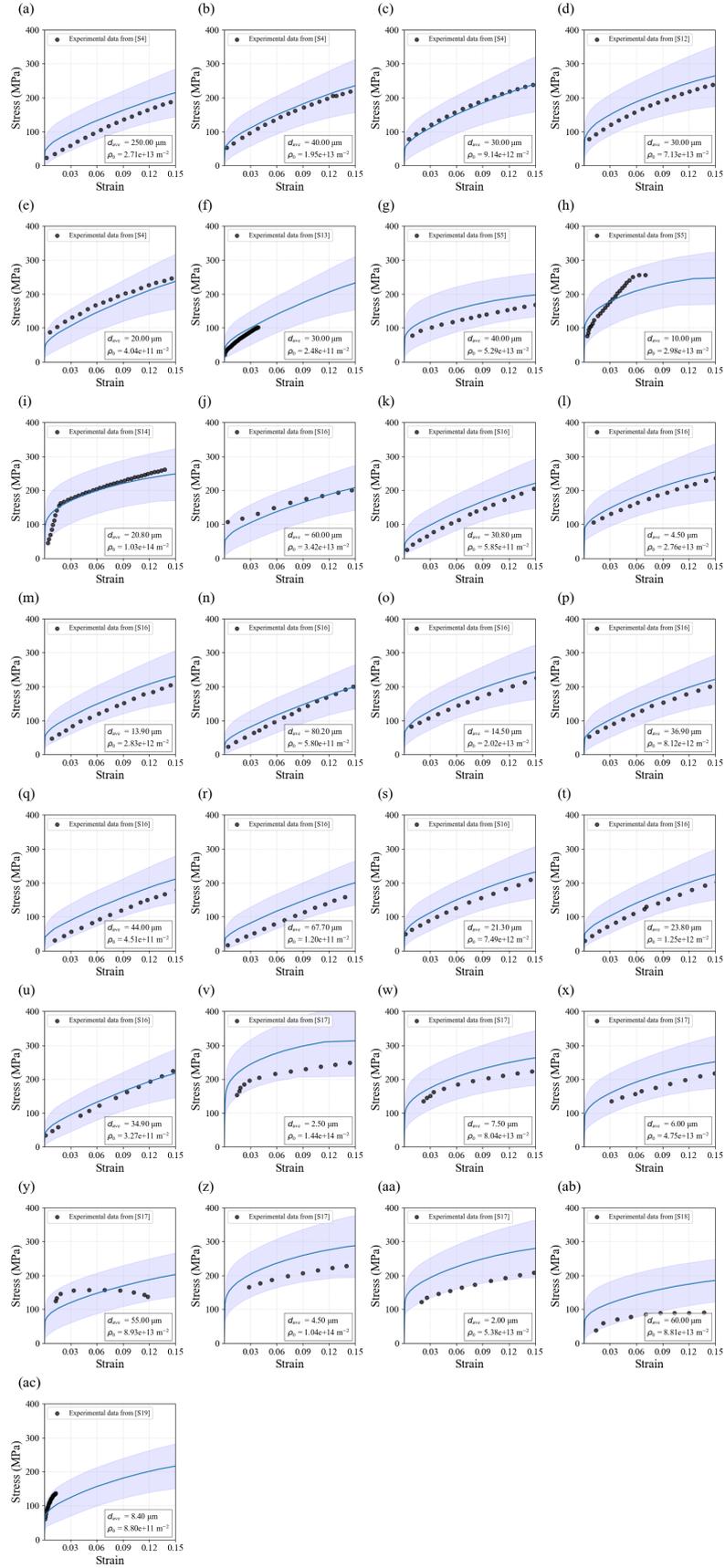

**Fig. S5** Comparisons of the predicted stress strain curves and the corresponding experiment for pure Cu collected from literature. In all figures the mean is shown by the solid line and the uncertainty range is shown by the shaded region. The corresponding experimental result is shown as solid circles.

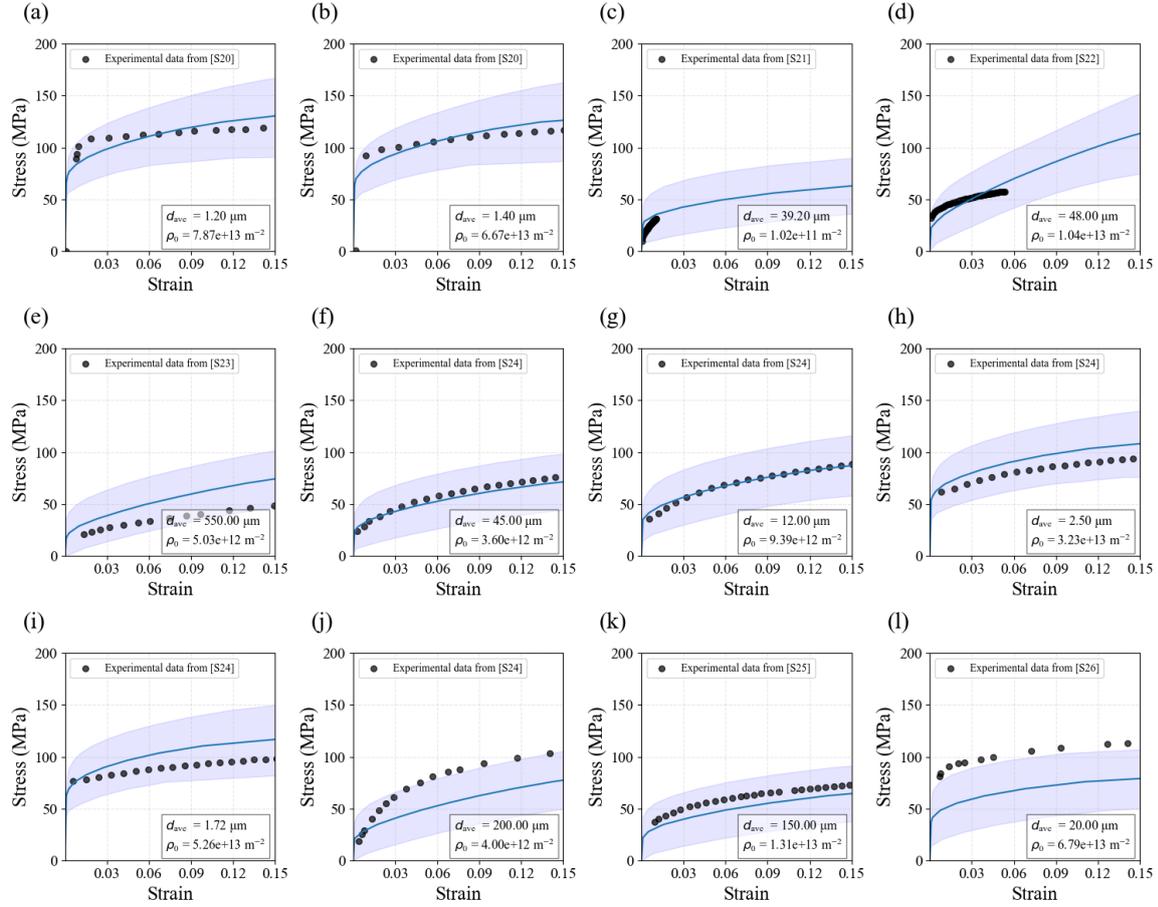

**Fig. S6** Comparisons of the predicted stress strain curves and the corresponding experiment for pure Al collected from literature. In all figures the mean is shown by the solid line and the uncertainty range is shown by the shaded region. The corresponding experimental result is shown as solid circles.

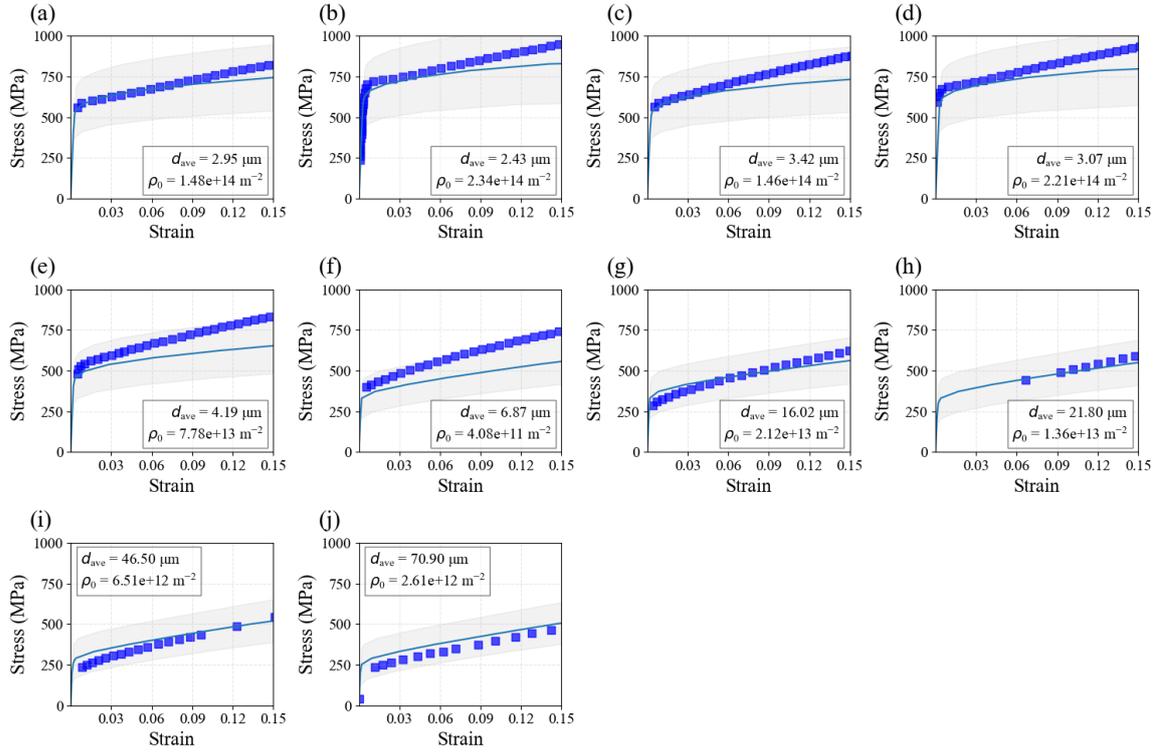

**Fig. S7** Comparisons of the predicted stress strain curves and the corresponding experiment for NiCoCrMnFe from this study. In all figures the mean is shown by the solid line and the uncertainty range is shown by the shaded region. The corresponding experimental result is shown as solid circles.

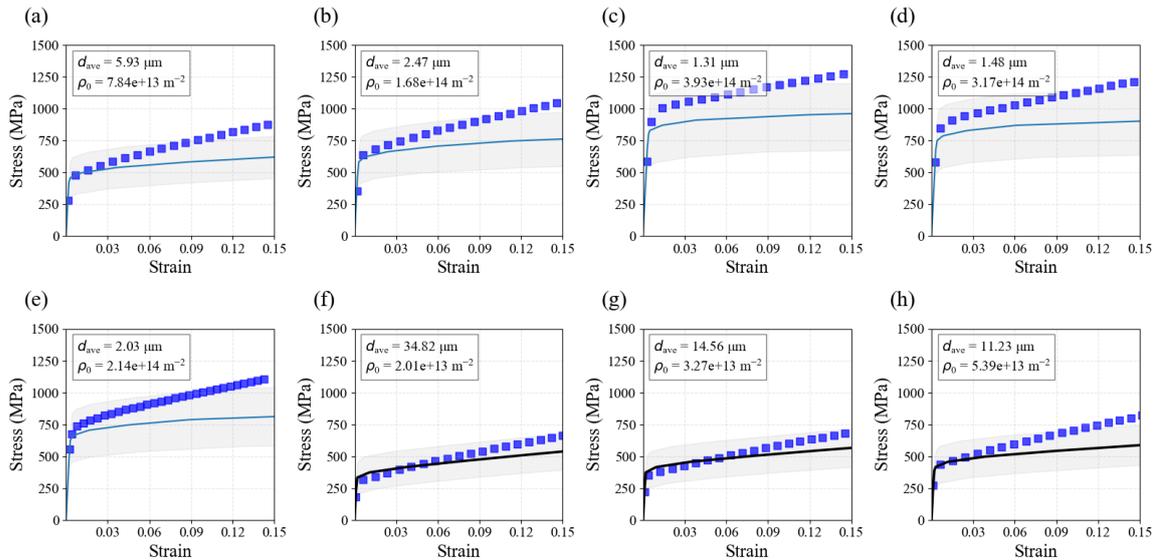

**Fig. S8** Comparisons of the predicted stress strain curves and the corresponding experiment for NiCoCr from this study. In all figures the mean is shown by the solid line and the uncertainty range is shown by the shaded region. The corresponding experimental result is shown as solid circles.

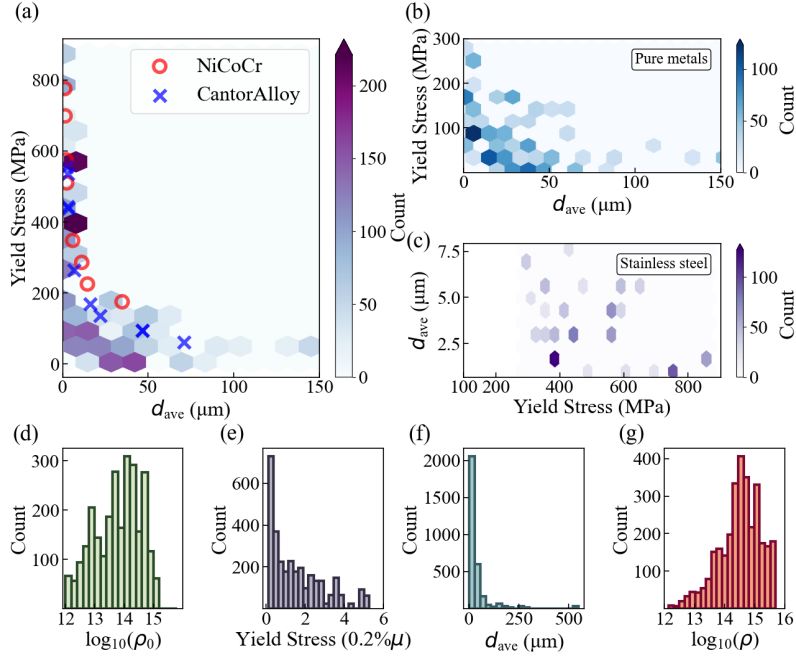

**Fig. S9 Summary of the collected experimental data set.** (a) Yield stress versus grain size for all data, with overlaid symbols indicate the new NiCoCr and NiCoCrMnFe data from the current study. (b) Yield stress versus grain size from the data set for pure FCC metals only (including Ni, Cu, and Al). (c) Yield stress versus grain size from the data set for the stainless steels data only (including 304L, 304, 316, 316L). (d) Distribution of the estimated $\rho_0$ for all experimental data. (e) Distribution of the $\rho$ from all stress-strain data. (f) Distribution of yield stresses. (g) Distribution of grain sizes.

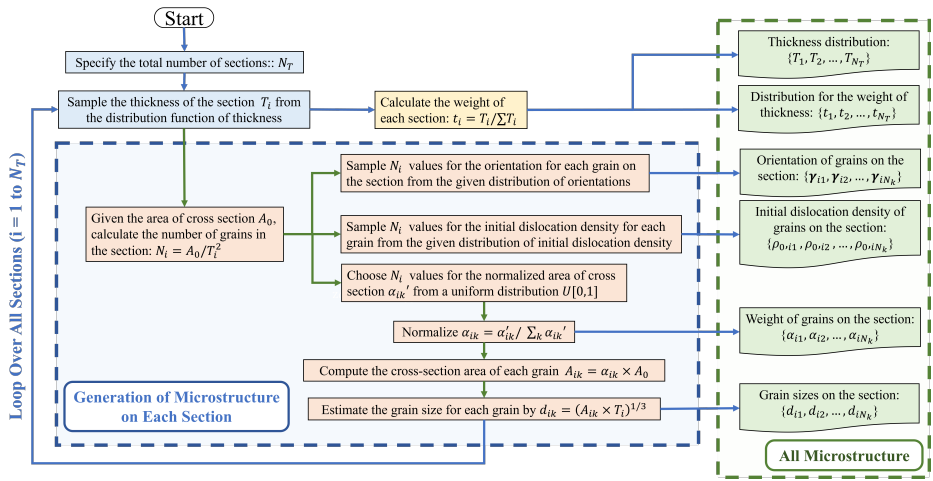

**Fig. S10** Flowchart for generating the synthetic 3D microstructures.

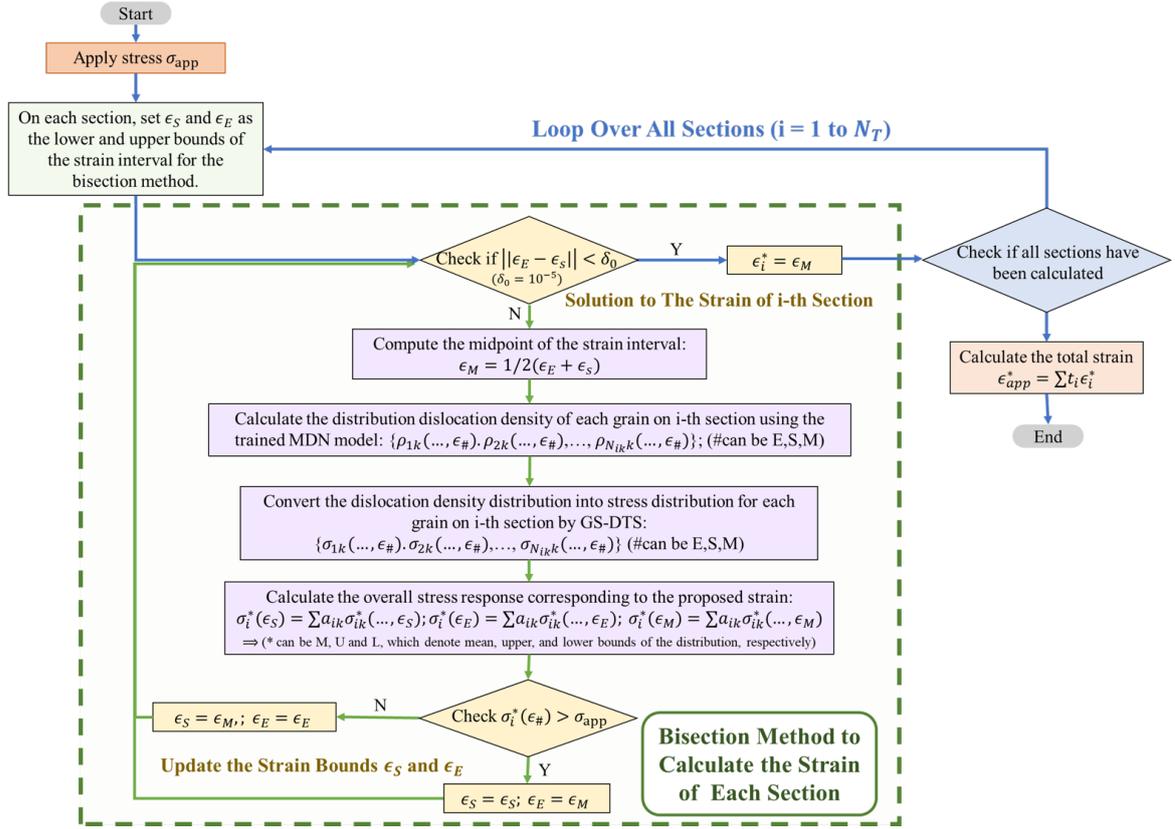

**Fig. S11** Flowchart for computing the strain corresponding to a prescribed macroscopic stress $\sigma_{\text{app}}$ using a bisection-based root-finding algorithm in the parallel–series polycrystal model.

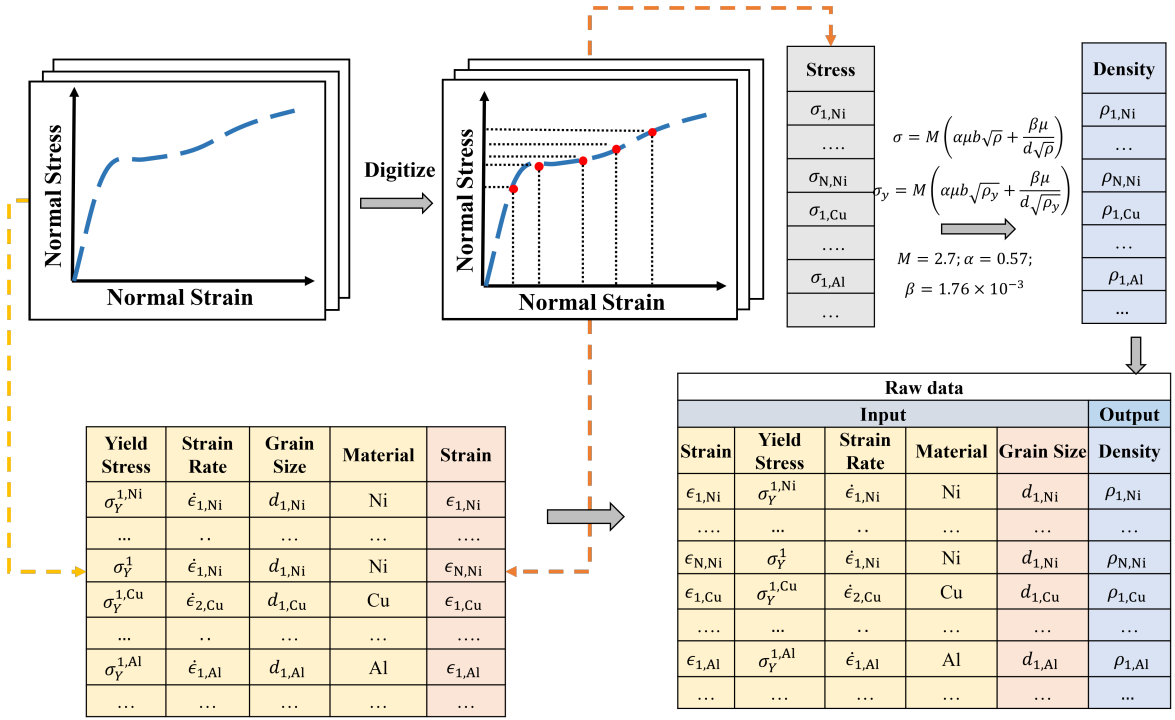

**Fig. S12** Overview of the data collection process. A stress–strain curve of Ni is digitized into discrete data points, and relevant information from the literature, such as strain values and material properties etc., is compiled into a structured input table. Corresponding stress values are converted to dislocation densities using the generalized size-dependent Taylor strengthening law. The result is a raw data set linking input features to dislocation density outputs.